\documentclass[conference,letterpaper]{IEEEtran}
\usepackage{graphicx} 
\usepackage{color} 
\usepackage{url} 
\graphicspath{{images/}}
\usepackage[cmex10]{amsmath}
\usepackage{amssymb}

\begin{document}

\title{Enabling Realistic Cross-Layer Analysis based on Satellite Physical Layer Traces}

\author{
\IEEEauthorblockN{Nicolas Kuhn$^{1,2}$, Emmanuel Lochin$^{1}$, J\'{e}r\^{o}me Lacan$^{1}$,  Roksana Boreli$^2$, Caroline Bes$^3$, Laurence Clarac$^3$}
\IEEEauthorblockA{\\
$^1$Universit\'{e} de Toulouse, ISAE, TeSA, Toulouse, France\\
$^2$NICTA Sydney, Australie\\
$^3$Centre National d'Etudes Spatiales, CNES, France\\
}
}

\maketitle

\begin{abstract}
We present a solution to evaluate the performance of transport protocols as a function of link layer reliability schemes (i.e. ARQ, FEC and Hybrid ARQ) applied to satellite physical layer traces. 
%We present a solution to evaluate the performance of transport protocols as a function of link layer reliability schemes (i.e. ARQ, FEC and Hybrid ARQ) applied on top of satellite physical layer traces. 
%The idea is to propose a realistic and practical solution to take into account physical layer traces inside a network simulator. The rationale is that modelling such traces might lead to approximation and errors and is sometimes difficult to build. Starting from a physical layer trace, we produce the equivalent link layer output to allow the use of this resulting trace inside a network simulator. 
As modelling such traces is complex and may require approximations, the use of real traces will minimise the potential for erroneous performance evaluations resulting from imperfect models. Our Trace Manager Tool (TMT) produces the corresponding link layer output, which is then used within the ns-2 network simulator via the additionally developed ns-2 interface module. 
%In order to demonstrate how our Trace Manager Tool (TMT) can be utilized to drive an efficient cross-layer analysis, we have developed a module that reads such output inside ns-2. We detail in this paper the analytical models built and then demonstrate that theoretical metrics (recovery delay and throughput efficiency) correctly match those obtained with TMT. 
We first present the analytical models for the link layer with bursty erasure packets and for the link layer reliability mechanisms with bursty erasures. Then, we present details of the TMT tool and our validation methodology, demonstrating that the selected performance metrics (recovery delay and throughput efficiency) exhibit a good match between the theoretical results and those obtained with TMT. Finally, we present results  showing the impact of different link layer reliability mechanisms on the performance of TCP Cubic transport layer protocol.
\end{abstract}

\section{Introduction}

The performance of satellite communications is mainly driven by the bit-error rate and the link delay. In order to improve the quality of video broadcasting or safety communications, existing interactions between reliability mechanisms at the link and other layers such as the transport layer must be considered. Data reliability can be operated independently at different levels of the communication stack (i.e. at every layers of the OSI model) and cross-layering techniques aim to optimize network usage while enabling better communications between individual layers. Reliability mechanisms exist at the transport layer (e.g. TCP enables an ARQ retransmission schemes), at the application layer (e.g. AL-FEC) or at lower layers such as the physical and link layers (e.g. ARQ, HARQ, erasure codes). In \cite{error_control}\cite{arq_schemes}, the authors present different mechanisms (FEC, ARQ and HARQ of type II) for the data transmission reliability but do not analyse their performance over a bursty channel. Using the results of \cite{markov_one_state}, the authors in \cite{stat_block_bursty} show that bursty errors can be modeled as a Markov chain.

While there are many studies that assess the impact of burst-correction codes on the physical layer level using this channel model \cite{fec_arq_burst}\cite{burst_code}\cite{effect_data_link}, little attention has been directed to the link layer where the channel impairments, seen as packet erasures, are approximated with a Bernoulli channel. As a result, the performance of erasure correcting codes over bursty erasure channel at the link layer, in terms of trade-off between throughput efficiency and recovery delay, has not been extensively studied. Especially in the context of Land-Mobile Satellite (LMS) channel and geostationary satellite systems, it is 
%obviously 
more realistic to consider bursts of losses at the link layer level, while it is acceptable to neglect the impact of queuing delays and processing times as these are close to negligible compared to the transmission delay. Therefore, one contribution of this paper is an analytical tool that enables 
%fast - NB RB: not sure what do you mean by fast, quick as in less than 3 seconds :-). Suggest to replace with efficcinet or remove
evaluations of the performance of different link layer reliability schemes. We propose to assess the impact of these link layer reliability mechanisms over a bursty erasure channel modelled by a Markov chain. The resulting algorithms enable us to evaluate the performance of reliability mechanisms over satellite links. 

To this end, we have developed a Trace Manager Tool (TMT) that computes the impact of a given reliability scheme at the link layer level as a function of a given physical layer trace. We combine TMT with the network simulator ns-2 to study the impact of link layer reliability mechanisms on the transport protocols. In~\cite{aimd}, the authors use a similar approach. However, they consider a simple hybrid space-terrestrial network and only focus on one link layer scheme (ARQ). %In this paper, we introduce a module for the network simulator ns-2 that exploits link layer trace to schedule the emission of IP packets. 
The authors in~\cite{ll_ns2} also consider link layer data inside ns-2. Compared to our proposal, they attempt to model 
the erasures and delay introduced by the reliability schemes at the link layer making their implementation inflexible and their 
results applicable to only a single physical layer model.
% results specific to only one physical layer model.
To the best of our knowledge, there is a clear lack of tools allowing the evaluation of all existing reliability schemes at the link layer following real physical measurements, and our contribution fills the gap in this area 
%and makes the link with transport protocols implemented inside a simulator such as ns-2. 
while utilising the rich existing source of transport protocol implementations within a simulator such as ns-2. 
%This paper is firstly dedicated to the implementation and validation of TMT (up to Section \ref{sec:cross_valid_and_interpretation}) and a preliminary cross-layer study is given in illustration in Section \ref{sec:ns2}. A future work will consider larger cross-layer studies based on the most recent transport protocols available in ns-2.

%We have chosen to organise this paper as follows: 
We have organised this paper as follows: in Section \ref{sec:bcm}, we present a bursty channel model and the method used to estimate erasure probabilities. We detail the algorithms used to evaluate the throughput efficiency and the recovery delay induced by these mechanisms in Section \ref{sec:algo}. In Section \ref{sec:cross_valid_and_interpretation}, we present the cross validation between the theoretical results and TMT. %We also illustrate in this section the use of TMT in a possible use case and analyse the results obtained. 
Section \ref{sec:ns2} presents the TMT tool and the module developed in ns-2 to introduce the reliability schemes at the link layer. This section also illustrates the performance of TMT based on a TCP Cubic example. Finally, Section \ref{sec:conc} presents the conclusion and future work.

\section{Link layer bursty erasure packet model}
\label{sec:bcm}

%This section explains how we transpose a bursty bit error channel (physical layer) model to derive a bursty erasure packets model (link layer). 
This section explains how we use the concept of a bursty bit error channel (physical layer) model to derive a bursty erasure packet (link layer) model. 
We base our analysis on the algorithms presented in \cite{canal_gilbert} to model link layer reliability schemes with a slight adaptation. In \cite{canal_gilbert}, the authors propose two methods to express the error probabilities of an error correcting code over a bursty channel. In particular, they provide a complete expression and computation method for the bit error probability (i.e. at the physical layer level). In our context, we need to modify these results by considering erasures at the link layer. 

\subsection{2-State Markov Chain}

%The model is defined as follows: the channel has two states and each state is characterized by an error probability and a changing state probability. 

%The 2-State Markov chain is illustrated in Fig.~\ref{markov_chain}.

%\begin{figure}[!ht]
%    \begin{center}
%	\includegraphics[width=0.9\linewidth]{good_bad.eps}
%	\caption{2-State Markov chain}
%	\label{markov_chain}
%    \end{center}
%\end{figure} 

%The good state probability (resp. bad state) presents an error probability $P_{G}$ (resp. $P_{B}$) and a changing state probability $\alpha$ (resp. $\beta$). In the good state (resp. bad state) errors occur with low (resp. high) probability, which illustrates the bursty aspect of the channel. With this definition of the channel, we %can define variable both burst lengths of lost packets ($\alpha$ and $\beta$) and errors occurrence in each state (with $P_{G}$ and $P_{B}$).
%can define two variables: the burst length of lost packets (respectively $\alpha$ and $\beta$) and the occurrence of errors in each state (with $P_{G}$ and $P_{B}$).

\begin{figure}[!ht]
    \begin{center}
	\includegraphics[width=1\linewidth]{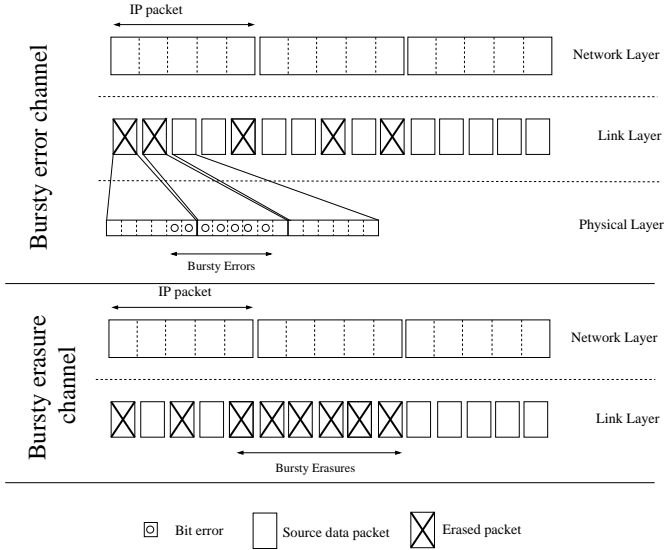}
	\caption{Bursty errors and bursty erasure models}
	\label{channel_model}
    \end{center}
\end{figure} 

A 2-State Markov chain model %Fig. \ref{markov_chain} 
is commonly used to represent a bursty error channel at the physical layer level. The good state probability (resp. bad state) presents an error probability $P_{G}$ (resp. $P_{B}$) and a changing state probability $\alpha$ (resp. $\beta$). In the good state (resp. bad state) errors occur with low (resp. high) probability, which illustrates the bursty aspect of the channel. We also use this model with corresponding erasure probabilities to simulate bursty erasures at the link layer level as illustrated in Fig.~\ref{channel_model}. In the context of satellite transmissions, this model if of interest as long bursts of erasures might occur.

\subsection{Evaluation of Erasure Probabilities}

%The recovering capacity of useful packets of an erasure code depends on the number of erasures over both useful and repair packets. 
For an erasure code, the capability to recover data packets depends on the number of erasures over both data and repair packets. The erasure probability distribution during a transmission over a channel with memory can be analysed through a 2-State Markov chain model, consisting of: the previous state of the chain, the probability to change the state and the corresponding erasure probability. As this 2-state Markov chain model applies to every packet, the totality of different erasure combinations over a number of packets can be considered through a mathematical induction.
% are estimated for each packet thanks to this 2-State Markov chain model. 
%Therefore, all the different repartitions of erasure over several packets can be considered through a mathematical induction.

We now present the iterative methods used in the following analysis. Let $P(m,n)$, be the probability of having $m$ erasures over $n$ packets, $P_{G}(m,n)$ (resp. $P_{B}(m,n)$) the probability to have $m$ erasures over $n$ packets and to be in the good state (resp. bad state) when the $n^{th}$ packet is received. In order to compute $P(m,n)$, we drive a double mathematical induction over $m$ and $n$, considering first the current state of the chain, and then the current erasure probability. We validate the implementation of this algorithm by comparing it's results with the results provided in \cite{canal_gilbert}.

\section{Modelling link layer reliability mechanisms with bursty erasures}
\label{sec:algo}

\begin{figure}[!ht]
  \begin{center}
		\includegraphics[width=1\columnwidth]{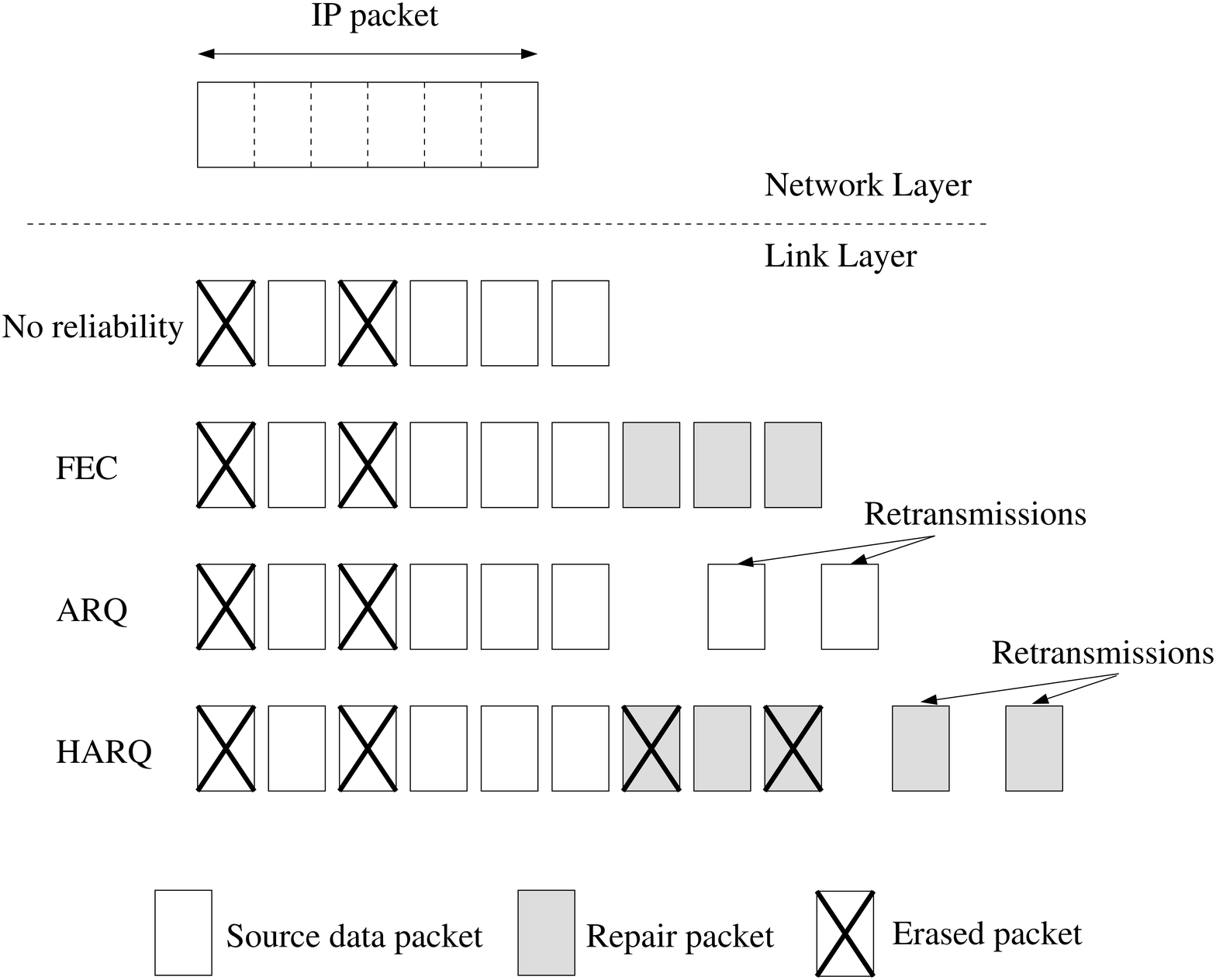}
		\caption{Error control mechanism}
		\label{error_control_mechanism}
  \end{center}
\end{figure}

We now present the expressions for the throughput efficiency and the recovery delay for specific reliability schemes (FEC, ARQ and HARQ of type II) at the link layer level. These mechanisms are presented through an example in Fig.~\ref{error_control_mechanism}. In the following, a full 'IP packet' is fragmented into Link Layer Data Units (denoted LLDU) before transmission over the physical layer. 

%Queuing delays, processing time or heavy traffic 
Queuing delays and processing times are considered in standard link layer models. As our analytical model is designed specifically for satellite links (with round trip times in the order of 400msec or greater), the impact of these additional delays can be neglected in comparison to the round trip time delay.

\subsection{FEC: Forward Error Correction}

In the FEC scheme, the sender sends a combination of data and repair LLDUs. Let $N_D$ (resp. $N_R$) be the number of data (resp. repair) LLDUs and $N=N_D+N_R$. The process to recover data LLDUs is successful if at least $N_D$ LLDUs are received, otherwise (if the number of erasures is strictly greater than $N_R$) no correction is possible.
First, we define the throughput efficiency as the ratio of the received LLDUs and the total number of LLDUs sent:

\begin{equation}
\eta_{FEC}=\frac{\sum_{i=1}^{N_D} P_{R}(i).i }{N_D+N_R}
\label{fec_eff}
\end{equation}

where $P_{R}(i)$ represents the probability that $i$ LLDU are received. Over a bursty erasure channel, this is computed following the previously explained mathematical induction: $P_{R}(i)=P(i,N_D+N_R)$ where $P(m,n)$ is the probability of having $m$ erasure in $n$ LLDUs.

Second, if a LLDU is erased, the additional delay will correspond to the time needed to receive the whole IP packet (data and repair LLDUs) needed by the FEC scheme to evaluate whether this IP packet can be recovered. 
This recovery delay, $d$, is related to the position of the LLDU in the total IP packet. On the average, we can consider the erasure to be located in the middle of the IP packet. With this hypothesis, the recovery delay $d$ for packets at the receiver can be calculated as: 

\begin{equation}
d=\frac{RTT}{2}+p \left ( 1-\sum_{i=N_R}^{N-1} P(i,N-1) \right ) \frac{N}{2} T_{P}
\label{fec_del}
\end{equation}

where $T_{P}$ is the time needed to receive a LLDU and $p$ the global erasure probability. 

\subsection{Interleaved FEC}

Interleaving is an efficient and commonly used technique to improve the data transmission over a bursty channel, as erasure bursts can be spread into a number of different codewords. It is possible to change the characteristics of the channel with (\ref{s2p}) in order to consider the interleaving. Let $p$ and $\rho$ be the local erasure probability and the correlation between the states (considering a simplified channel with $P_G=0$ and $P_B=1$). We have:

\begin{equation}
	p=\frac{1-\alpha}{2-\alpha-\beta} \text{~~and~~} \rho=\alpha+\beta-1
	\label{s2p}
\end{equation}

If an interleaver with a depth $I$ is used on this bursty channel, we obtain a new bursty channel with the following changing state probabilities $\alpha_I$ and $\beta_I$:

\begin{equation*}
	\alpha_{I}=p+\rho^{I}.(1-p)  \text{~~and~~} \beta_{I}=(1-p)+\rho^{I}.p
\end{equation*}

The performance of interleaved FEC can be then obtained by applying parameters $\alpha_I$ and $\beta_I$ in equations (\ref{fec_eff}) and (\ref{fec_del}).

\subsection{ARQ: Automatic Repeat-reQuest}

Automatic Repeat-reQuest mechanism at the link layer level consists in the retransmission of the LLDU that have been lost during the transmission. The throughput efficiency (also called goodput which is, by definition, the application layer throughput) corresponds to the probability that a LLDU is received. In the context of high delay links, the channel probably changes its state before retransmissions are sent. Thus, we do not consider burst of erasures when using ARQ. Furthermore, we can neglect this notion as this scheme does not introduce correlation between different LLDU of the same IP packets. Then, the recovery delay can be expressed as follows:

\begin{equation*}
d_{ARQ}=\frac{RTT}{2} +\sum_{i=1}^{\infty} p^{i-1}(1-p) i.RTT
\end{equation*}
where $p$ is the global erasure probability.

\subsection{HARQ-II: Hybrid ARQ of type II}

This mechanism is a combination of the FEC and ARQ mechanisms and after the first transmission of a FEC block, including data and repair LLDUs, HARQ-II allows the sender to send additional repair LLDUs when a recovery is not possible at the receiver side. In other words, if no correction is possible at the receiver, the transmission of additional repair LLDUs is requested by the receiver. At each new transmission, the sender transmits more LLDUs than requested by the receiver: if the receiver requires $n$ LLDUs to recover the data, the transmitter sends $(n+N_S)$ LLDUs when $N_S$ is the number of supplementary repair LLDUs sent. Let $\text{R}_{r}$ be the probability that the data can be decoded after $r$ retransmissions, $T_{R}(r)$ the time needed to receive the LLDU of the $r^{th}$ retransmission, $N_D$ the number of data source LLDUs, $N_R$ the number of repair LLDUs, and $N=N_D+N_R$. For applications with time constraints, a limited number of authorized retransmissions, denoted by $R$, is considered.

\subsubsection{Throughput Efficiency}

The throughput efficiency for HARQ-II is expressed as the ratio of the received LLDUs and the total number of LLDUs sent:  

\begin{equation*}
\eta_{HARQ}=\frac{\sum_{i=1}^{N_D} P_{R}(i).i }{\sum_{j=1}^{\infty} P_{S}(j).j}
\end{equation*}

where $P_{R}(i)$ is the probability that $i$ LLDU are received and $P_{S}(j)$ the probability that $j$ LLDU are sent. 

\begin{equation}
\sum_{i=1}^{N_D} P_{R}(i).i= \left ( \sum_{z=0}^{R-1} R_{z} \right ) .N_D + \sum_{i=1}^{N_D-1} P_{R}(i).i 
\label{Pi}
\end{equation}

with $R=2$ (2 complementary transmissions are authorized), (\ref{Pi}) can be calculated according to the following expression:

\begin{equation*}
\begin{aligned}
& P_{R}(i, i<N_D )=P(N_D-i,N_D)\\
& \times \sum_{l_{1}=\delta_{i,N_D,N_R}}^{N_R} \sum_{l_{2}=N_S+1}^{(N_D-i)+N_S-N_R+l_{1}} \sum_{l_{3}=N_S+1}^{l_{2}+N_S} \Delta_{i,N_D,N_R,N_S,l_{1},l_{2},l_{3}}  
\end{aligned}
\end{equation*}

with:

\begin{equation*}
\begin{aligned}
&\Delta_{i,N_D,N_R,N_S,l_{1},l_{2},l_{3}} =P(l_{1},N_R)\\
& \times P(l_{2},(N_D-i)+N_S-N_R+l_{1})\\
& \times P(l_{3},l_{2}+N_S)
\end{aligned}
\end{equation*}

and

\begin{equation*}
\delta_{i,N_D,N_R}=
\begin{cases}
& 0 \text{\ if } (N_D-i)>N_R \\
& (N_D-i)-N_R \text{\ if } (N_D-i)<N_R\\
& 1 \text{\ if } (N_D-i)=N_R
\end{cases}
\end{equation*}

We consider every combination of erasure positions to determine $P_{S}(j)$. For each complementary transmission, the number of repair LLDU sent is linked to the current number of erasures. If there are $n$ erasures at the first IP packet (data and repair LLDU) sent, and if the correction capacity of the code is $N_R$, there are two possibilities: if $n \leqslant N_R$, no transmission of repair LLDU is needed; if $n>N_R$ the receiver requests for $n-N_R+N_S$ repair LLDU. The expressions (\ref{Pj}) are given with $R=2$ and with $P(m,n)$, the probability to have $m$ erasures over $n$ LLDU.

\begin{equation}
\begin{aligned}
 &P_{S}(j)=\sum_{l_{0}=0}^{N_R} \delta_{N}.P(l_{0},N) \\ 
 &+\sum_{l_{0}=N_R+1}^{N}\sum_{l_{1}=0}^{N_S} \delta_{l_{0},N}.P(l_{0},N).P(l_{1},l_{0}-N_R+N_S) \\ 
 &+\sum_{l_{0}=N_R+1}^{N}\sum_{l_{1}=N_S+1}^{l_{0}+N_S} \delta_{l_{0},l_{1},N}.P(l_{0},N).P(l_{1},l_{0}-N_R+N_S)
\end{aligned}
\label{Pj}
\end{equation}

with:

\begin{equation*}
\begin{cases}
& \delta_{N}= \begin{cases}
& 1 \text{\ if } j=N \\
& 0 \text{\ if } j \neq N
\end{cases} \\
& \delta_{l_{0},N}= \begin{cases}
& 1 \text{\ if } j=N+(l_{0}-N_R+N_S)\\
& 0 \text{\ if } j \neq N+(l_{0}-N_R+N_S)
\end{cases} \\
& \delta_{l_{0},l_{1},N}= \begin{cases}
& 1 \text{\ if } j=N+(l_{0}-N_R+N_S)+l_{1}\\ 
& 0 \text{\ if } j \neq N+(l_{0}-N_R+N_S)+l_{1}
\end{cases}
\end{cases}
\end{equation*}

\subsubsection{Recovery Delay}

We have to consider both the time needed to receive the first FEC block (data and repair LLDU), $T_{R}(0)$, and the additional repair LLDU.
This recovery delay, denoted $d_{HARQ}$ can be expressed as follows:

\begin{equation*}
d_{HARQ}=T_{R}(0)+\frac{RTT}{2}+\sum_{i=1}^{\infty} R_{i}.i (RTT+T_{R}(i))
\end{equation*}

with $RTT \gg T_{R}(i)$.

\subsection{Conclusion}

The theoretical expressions developed in this section will be used:
\begin{itemize}
\item to validate the TMT tool presented in the following section;
\item to quickly estimate the performance of the link layer reliability schemes.
\end{itemize}

\section{Cross-validation and interpretation}
\label{sec:cross_valid_and_interpretation}

In this section, we present the TMT tool and then measure the resulting throughput efficiency and recovery delay over link layer output. 
We then cross-validate the TMT tool results with the theoretical metrics previously in Section~\ref{sec:algo}. 

\subsection{The Trace Manager Tool: TMT}

We present the Trace Manager Tool that implements the standard link layer reliability schemes (i.e. ARQ, FEC, H-ARQ). %The main idea of this software is to produce the output of the link layer based on physical layer trace.

The input data of TMT consists in a list of parameters (reliability scheme used, $N_D$, $N_R$, $RTT$, etc.) and the physical layer trace considered. We propose two ways to use the physical layer input, depending on the origin of the erasures:
\begin{itemize}
\item direct use of the physical layer trace: the physical traces are measured and erasure events occur at the link layer according to real channel evolutions and to the most recent codes at the physcial layer;
\item indirect use of the physical layer trace: erasures are introduced on one error-free input trace following a 2-State Markov chain model as explained in Section~\ref{sec:bcm}.
\end{itemize}

In the use case presented, the physical trace corresponds to a satellite data transmission with a duration of of 500 seconds and has been provided by courtesy of CNES\footnote{CNES is a government agency responsible for shaping and implementing France's space policy in Europe, see \url{http://www.cnes.fr/}.}. As the physical trace provided is error-free, we thus introduce bursty erasures over this physical layer trace following the Markov chain model presented in Section~\ref{sec:bcm}.
TMT computes the equivalent output link layer trace according to the input trace and the chosen reliability scheme. We only keep useful data LLDU (and not repair LLDU or retransmission) and adapt their decoding delay according to the reliability scheme chosen. Then, we compute the throughput efficiency (i.e. goodput) and the recovery delay.

The basic principle of TMT is to map available LLDU at the link layer to incoming IP packets and to schedule the emission of the packets at the transport layer level conjointly with the reliability schemes introduced at the link layer level. Later in Section~\ref{sec:ns2}, we illustrate how TMT allows to assess the impact of link layer reliability schemes on transport layer protocols performance.

\subsection{Validation}

\begin{figure}[htb]
\begin{center}
\includegraphics[width=0.67\columnwidth]{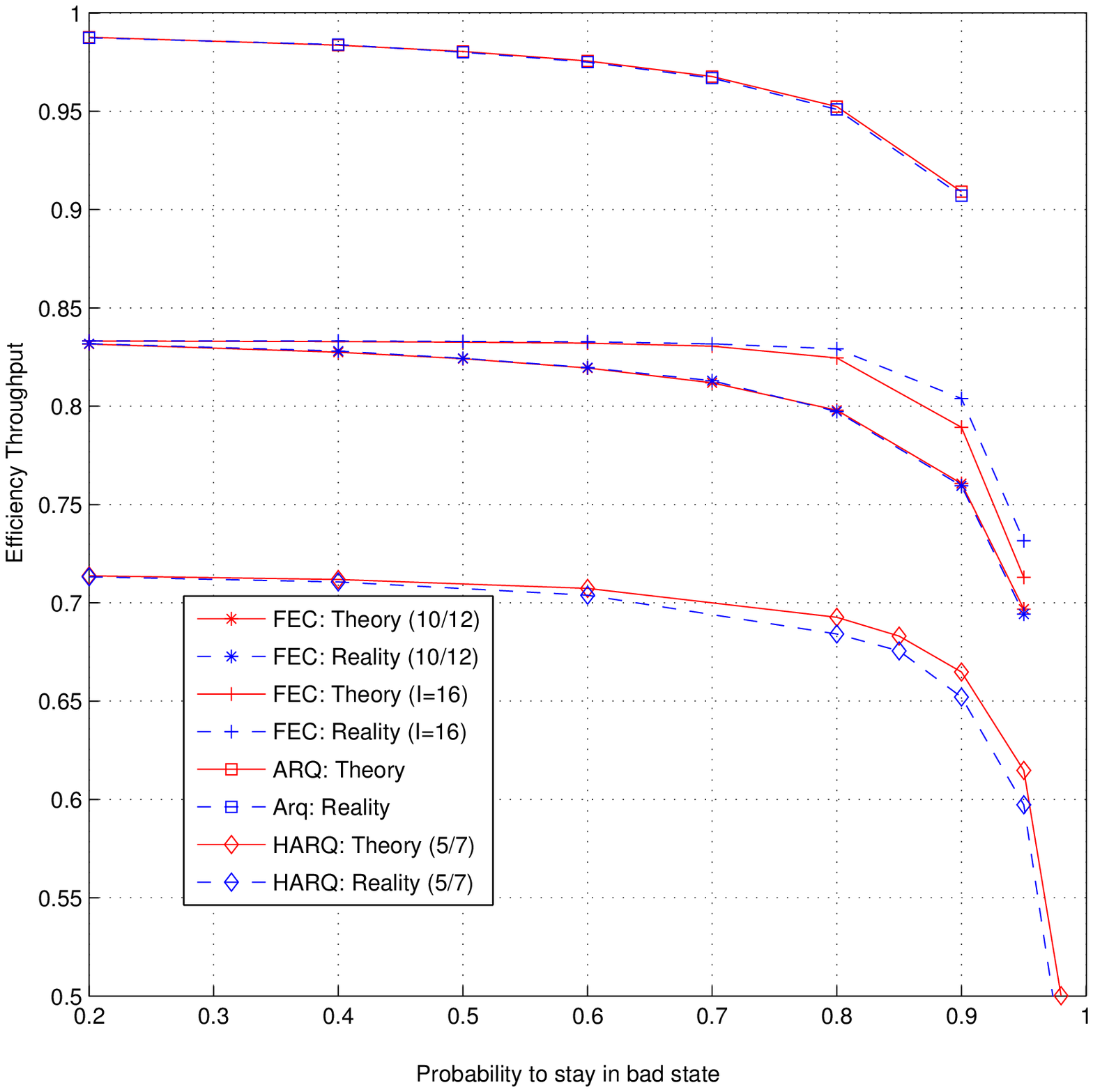} 
\includegraphics[width=0.8\columnwidth]{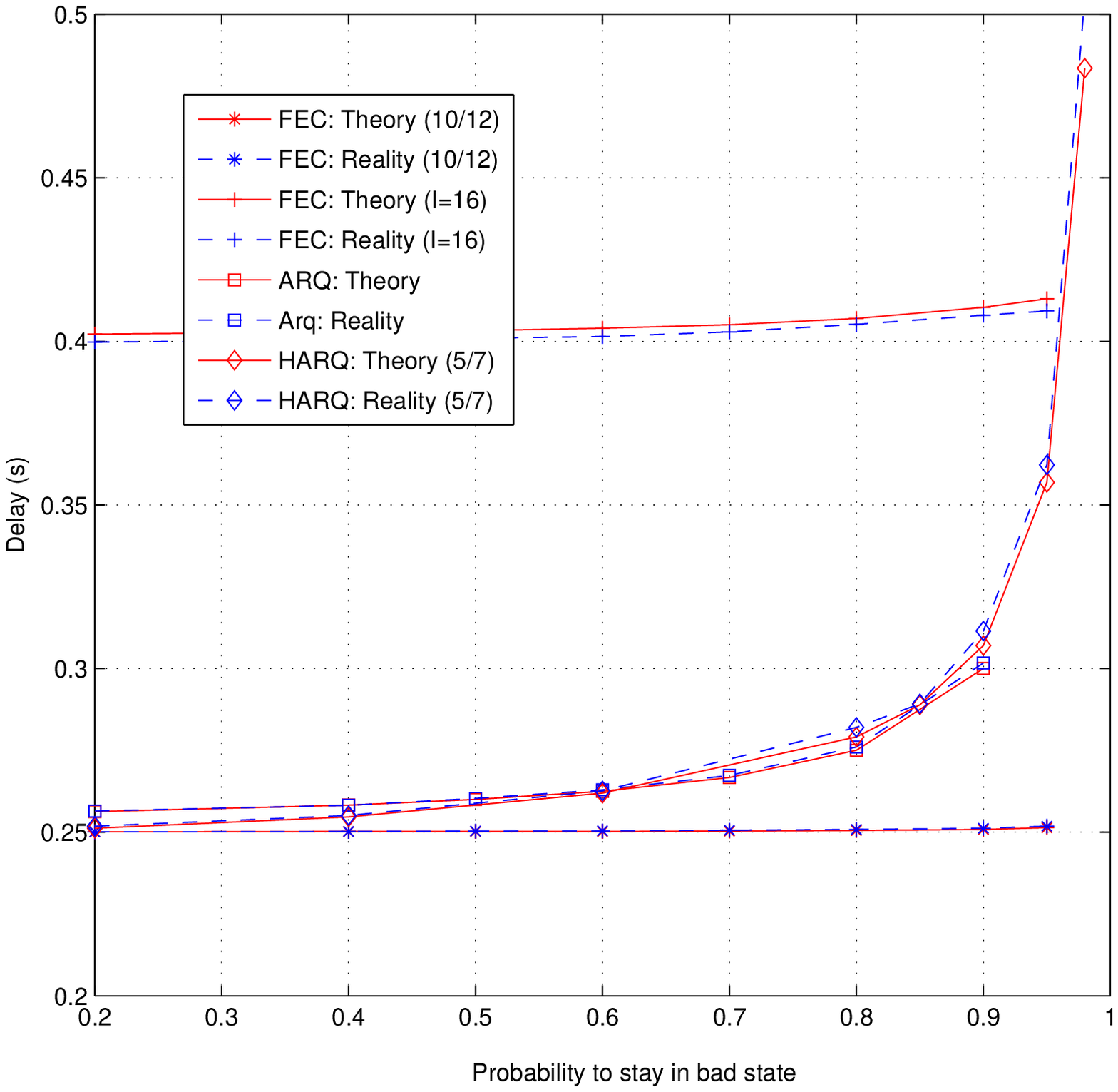}
\caption{Validation of the throughput efficiency and the recovery delay}
\label{val_eff_del}
\end{center}
\end{figure}

For each state, we compute the theoretical metrics through the equations detailed in Section~\ref{sec:algo} and the resulting metrics obtained with TMT. We present in Fig.~\ref{val_eff_del} the results obtained on a given set of parameters. The chosen parameters are: $RTT=500ms$, $N_{D-FEC}=10$, $N_{R-FEC}=12$, $N_{D-HARQ}=5$, $N_{R-HARQ}=7$, $\alpha=0.99$, $\beta \in [0.1;0.98]$, which induced a global erasure probability $p \in [0.01;0.3]$ and a length of erasure bursts $t_{b} \in [1;50]$. Both figures confirm that the theoretical expressions developed fits TMT results. Note that we only present a subset of our experiments and that several other set of parameters have been tested with success. %We thus consider that both tools are cross-validated.

\subsection{Interpretation}

We propose to exploit the theoretical expressions given in Section~\ref{sec:algo} to compare the three recovery mechanisms in terms of recovery delay and throughput efficiency over a bursty channel. For the simulation, we use the following parameters: $RTT=500ms$, $N_D=38$, $N_R=13$, $R=2$, $\alpha=0.98$, $\beta \in [0.1;0.93]$, which induces a global erasure probability $p \in [0.01;0.3]$ and a length of erasure bursts $t_{b} \in [1;14]$. Please note that the interpretation of the following results is limited to the given parameters.

\begin{figure}[htb]
\begin{center}
\includegraphics[width=0.8\columnwidth]{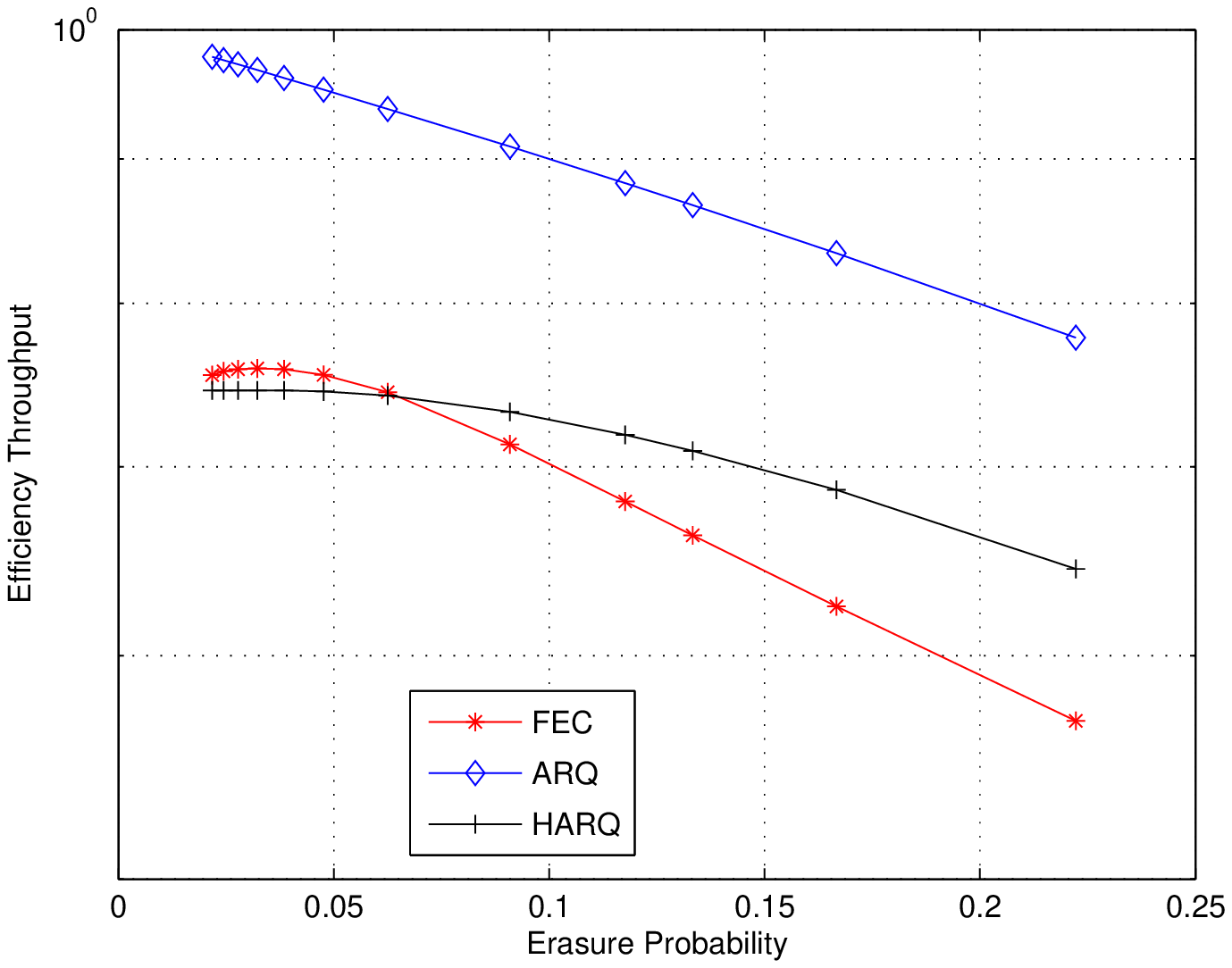}
\includegraphics[width=0.8\columnwidth]{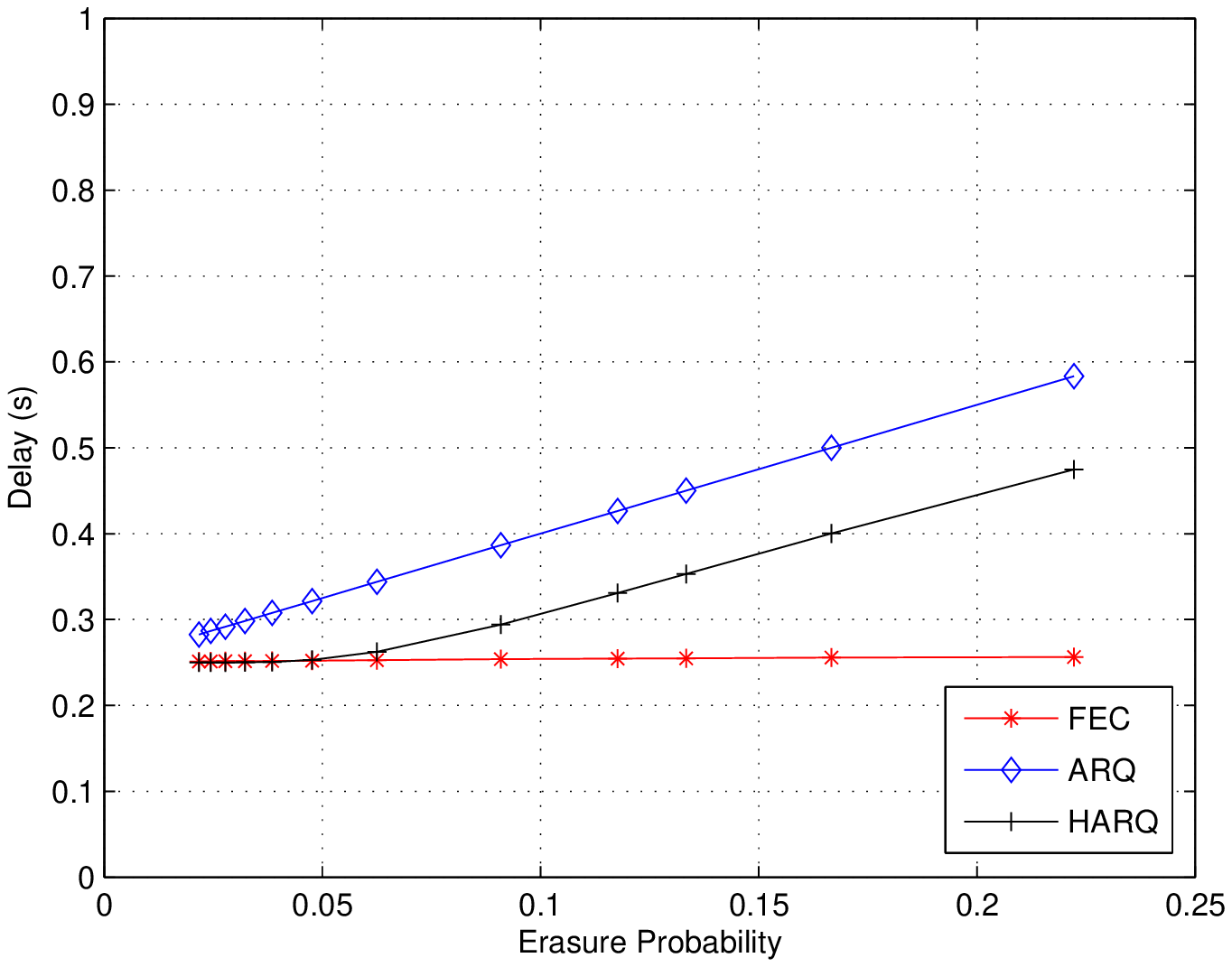}
\caption{throughput efficiency and recovery delay: use case}
\label{fig_eff_del}
\end{center}
\end{figure}

The results presented in Fig.~\ref{fig_eff_del} have been obtained with MATLAB. We note that ARQ and HARQ can transmit supplementary LLDUs if the IP packet cannot be rebuilt. %In the context of satellite links, the delay resulting from the retransmissions impacts the data delivery. 
In the context of satellite links, the delay resulting from the retransmissions impacts the data delivery and although these retransmissions enable the recovery of lost LLDUs at a later time, they may not be gainfully utilized by the time constrained applications. When erasure occurrence is low, ARQ demonstrates better performance than HARQ as the transmitter does not send useless repair LLDU. Therefore, when erasure occurrence is higher, HARQ introduces less delay thanks to the initial repair LLDUs. Although the transmission can be reliable with both ARQ and HARQ schemes, the introduced delay needs to be considered in the design of networks with time constraints. 
The theoretical models presented in this paper allow a fast analysis of the performance of reliability schemes over various channels and can consequently assist the network designer with the choice of the most appropriate scheme to be used.
%As a result, the theoretical models allow a fast analysis of the performance of reliability schemes over various channels. %This point will be further developed in a future work.

\section{Practical used of TMT with ns-2}
\label{sec:ns2}

%\subsection{Linkage between ns-2 and TMT}
%
%We now explain how the ns-2 module exploits link layer trace. We introduce a new queuing model that inherit from "Drop Tail": it enables us to keep the FIFO mechanism and the queue over-flow control. The link layer trace are loaded and the queuing mechanism use the arrival and decoding times of each link layer data unit to evaluate the date the "IP packet" should be sent at. The Fig.~\ref{fig_software_ns2} illustrate this based on an example. 
%\begin{itemize}
%\item The first "IP packet", denoted '1', is divided into 3 LLDU. We link the output date of this "IP packet" to the output date of the last of these 3 LLDU. In this example, the third LLDU has the greatest output time: the last bit of this "IP packet" is emitted when the emission of this LLDU is over. 
%\item The case of the "IP packet" denoted 2 illustrate that if one LLDU is missing, the whole "IP packet" is lost. 
%\item Our module consider scheduling problems induced by the various decoding times introduced by the reliability schemes at the link layer, as illustrated by the "IP packet" denoted 3 and 4. The network simulator ns-2 can not emit several packets at the same time, so we introduce a small delay and reorganize the queuing process. 
%\end{itemize}
%
%\begin{figure}[!h]
%  \begin{center}
%		\includegraphics[width=1\columnwidth]{ns2_crosslayer.eps}
%		\caption{ns-2 module}
		\label{fig_software_ns2}
%  \end{center}
%\end{figure}
%
%\subsection{An example of simulation}

We now present how to play TMT resulting output trace with the network simulator ns-2. We first generate link layer trace with TMT following the studied physical layer trace. The ns-2 simulator allows to drive simulation based on external traces. As a result, ns-2 loads the link layer trace and the standard ns-2 queuing mechanism uses the arrival and decoding times of each link layer data unit to evaluate the IP packet sending time. 
In this simulation, we consider a simple link between a satellite and a gateway. The transport layer uses TCP Cubic protocol. We aim to observe the performance of this transport protocol in terms of number of packets sent and retransmissions (at both link and transport layers) over different link layer reliability schemes. 

We use the same error-free physical trace previously introduced in Section \ref{sec:cross_valid_and_interpretation}. The objective of the study is to estimate the impact of reliability mechanisms when different losses distribution are introduced over this channel. We thus introduce bursty erasures over the physical layer trace following the Markov chain model presented in Section~\ref{sec:bcm}: the mean burst length corresponds to 12.5 link layer data units and the mean erasure probability is 20\% ($\alpha=0.98$ and $\beta=0.92$). We propose to focus on the ARQ and HARQ mechanisms ($N_{D-HARQ}=10$, $N_{R-HARQ}=20$).

%\subsection{Presentation of the results} 

\begin{figure}[!h]
\begin{center}
\includegraphics[width=0.8\columnwidth]{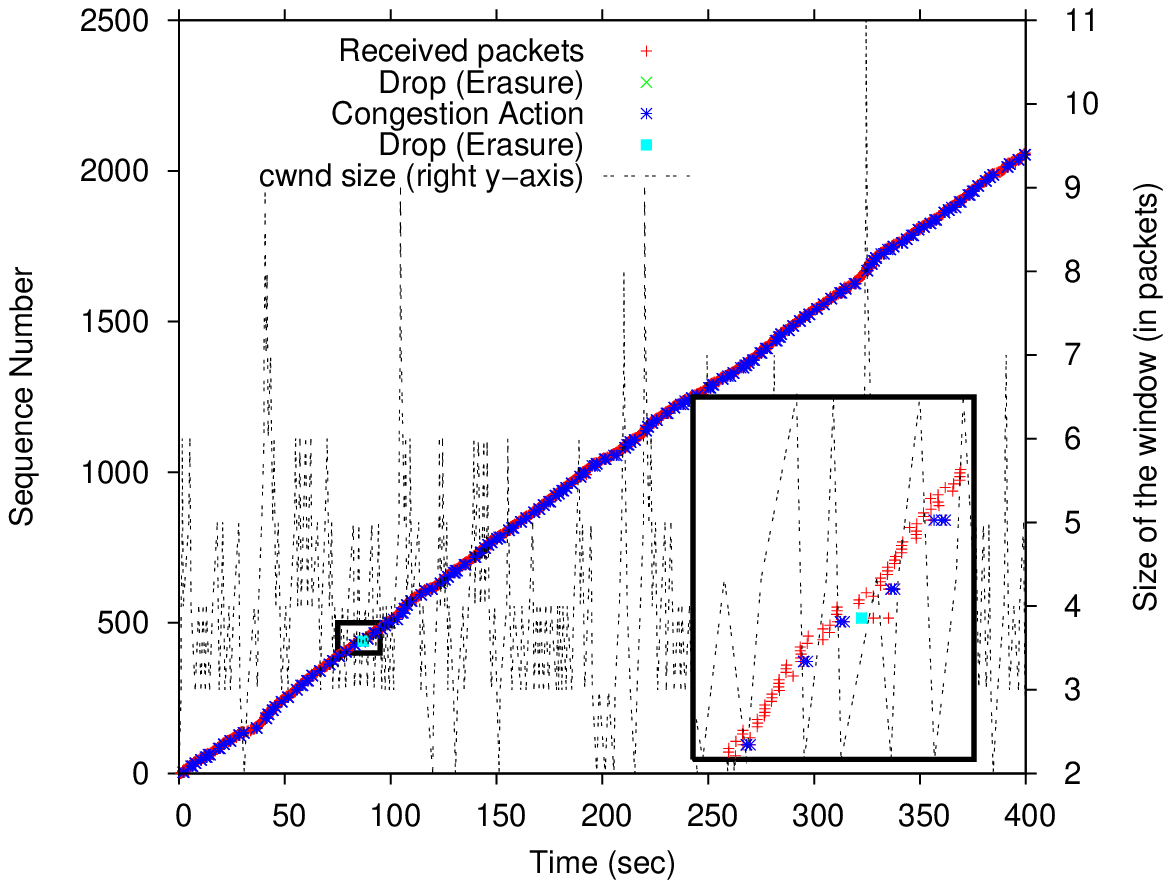}
\includegraphics[width=0.8\columnwidth]{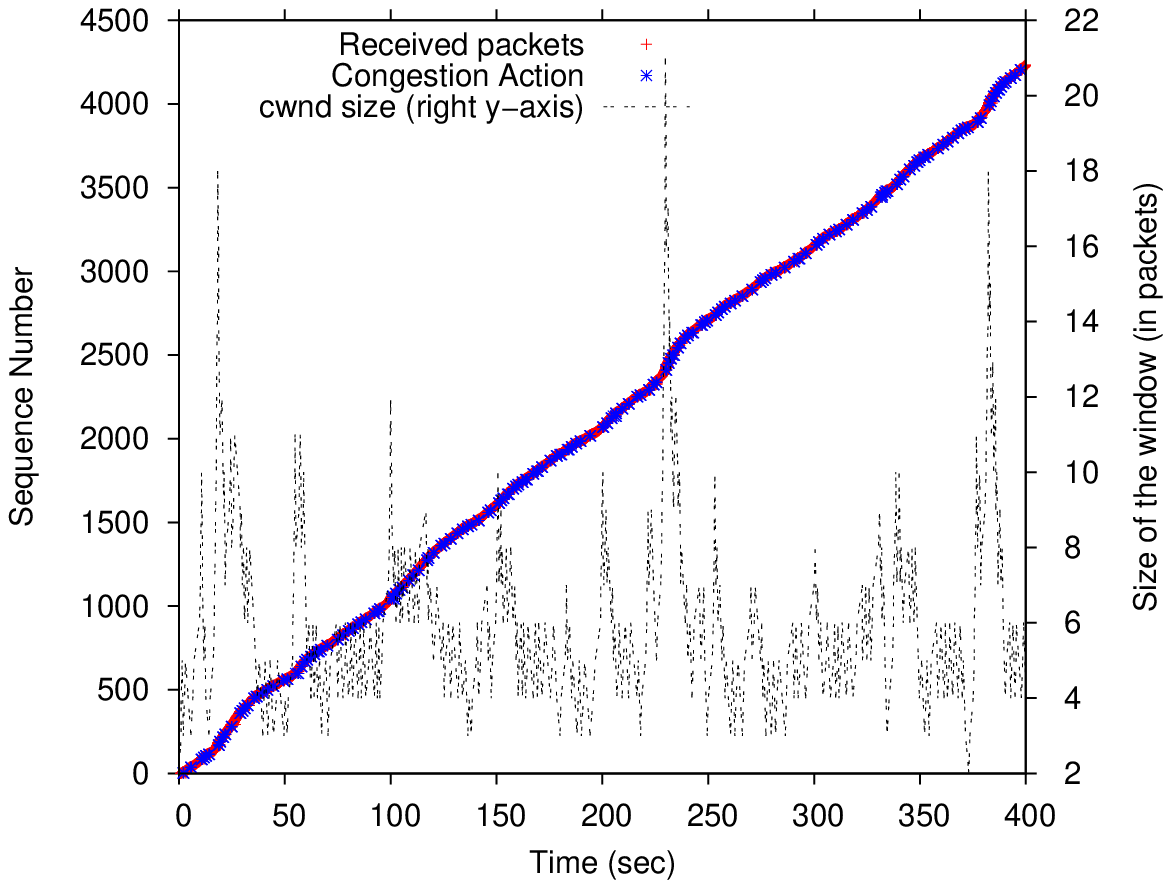}
\caption{Comparaison between ARQ and HARQ at the link layer}
\label{arq_goodput}
\end{center}
\end{figure}

\begin{table}[!h]
\begin{center}
\begin{tabular}[h]{|c|c|c|c|c|}
\hline
  & \multicolumn{2}{c|}{Transport Layer} & \multicolumn{2}{c|}{Link Layer} \\ \cline{2-5}
Number of transmissions & HARQ & ARQ & HARQ & ARQ \\ \hline
1 & 0.935 & 0.902 & 0.923 & 0.799 \\ \hline
2 & 0.064 & 0.093 & 0.060 & 0.159 \\ \hline
3 & 0.0009 & 0.005 & 0.013 & 0.033 \\ \hline
4 & 0 & 0 & 0.003 & 0.006 \\ \hline
5 & 0 & 0 & 0.0004 & 0.0012 \\ \hline
6 & 0 & 0 & 0.0001 & 0.0002 \\ 
\hline
\end{tabular}
\caption{Comparison of the probability of having a specific number of (re)transmissions for ARQ and HARQ}
\label{tab:comp_ret}
\end{center}
\end{table}

We sum up a sample of the results obtained in Fig.\ref{arq_goodput} and Table.\ref{tab:comp_ret}. Please note that we do not set the parameters to optimize the transmission. The objective is only to illustrate the capability of TMT. This example shows that with the same transport protocol, HARQ of type II at the link layer sends more packets than ARQ. As shown in Fig.\ref{arq_goodput}, we observe that with ARQ scheme (resp. HARQ of type II), the receiver acknowledged 2000 IP packets (resp. 4100 IP packets). As the number of retransmissions at the link layer is limited, the zoom square in Fig.\ref{arq_goodput} shows that an IP packet is dropped and then retransmitted. For both reliability schemes, 
%useless packets 
packets which are not useful are sent (redundancy packets for HARQ, and lost packets for both schemes) but HARQ introduces less delay (i.e. less retransmissions at the link layer level). Through this example, we illustrate that in case of noisy channels, HARQ-II outperforms ARQ. Moreover, as fewer packets are dropped at the transport layer, the retransmissions at this layer level are mainly spurious: we can perform further works with TMT with the aim to reduce the number of spurious retransmissions and 
%useless 
unnecessary congestion window decreases in the context of satellite links.

%As a matter of fact, the results obtained with TMT are of interest to clearly understand the impact of reliability schemes at the link layer on upper layers. 
We note that the results obtained with TMT directly assist with understanding the impact of reliability schemes at the link layer on the performance of upper layers. To the best of our knowledge, TMT is the first tool enabling such studies based on real physical layer traces.

%This subsection aims to present an example of the results we can obtain through our tool, without getting interested in the optimization of the reliability schemes. Indeed, we aim to obtain realistic physical trace from a software of the CNES and lead larger studies in the satellite context with the work we present in this paper in a future work.

\section{Conclusion}
\label{sec:conc}
In this article, we present a Trace Manager Tool (TMT) that computes the equivalent link layer output of a real physical trace as a function of the reliability schemes used (FEC, ARQ, HARQ). We propose a module for ns-2 that takes into account this link layer trace in order to study the impact of link layer reliability schemes on the performance of transport protocols. We provide theoretical expressions for the throughput efficiency and the recovery delay for these reliability schemes over bursty channels, enabling us to validate our TMT tool and to drive fast evaluations of their performance. The resulting model allows better assessment of the benefits brought out by the reliability mechanisms in terms of QoS for the applications. 

In future work, we plan to further utilise TMT and the ns-2 module in order to drive an extensive study of the impact of link layer reliability schemes on transport protocols performance, in the context of hybrid space-terrestrial networks.

\section*{Acknowledgement}
The authors wish to thank the CNES for their support.

\end{document}